\documentclass{elsart}

\usepackage{graphics}
\usepackage{graphicx}
\usepackage{epsfig}

\usepackage{amssymb}

\newcommand{\vp}{{\bf p}}

\newcommand{\bbf}{}

\newcommand{\bc}{\begin{center}}
\newcommand{\ec}{\end{center}}

\newcommand{\be}{\begin{equation}}
\newcommand{\ee}{\end{equation}}
\newcommand{\bea}{\begin{eqnarray}}
\newcommand{\eea}{\end{eqnarray}}
 \newcommand{\cal}{}
\begin{document}

\begin{frontmatter}



\title{Inelastic scattering from quantum impurities}


\author{Gergely Zar\'and and L\'aszl\'o Borda}

\address{ Department of Theoretical Physics, Budapest University of Technology and Economics,
Budafoki \'ut 8.  H-1521 Hungary}

\begin{abstract}
We review  the non-perturbative   theoretical framework set up recently to compute 
the inelastic scattering cross section from quantum impurities [G. Zar\'and {\it et al.},
Phys. Rev. Lett. {\bf 93}, 107204 (2004)] and show how it can be applied
 to a number of quantum impurity models. We first use this method for the 
$S=1/2$ single-channel Kondo model and the Anderson model. In both cases, a
large plateau is found in the inelastic scattering rate for incoming energies
above $T_K$, and a quasi-linear regime appears in the energy range  
$0.05\;T_K < \omega < 0.5\; T_K$, in agreement with the experimental observations. 
We also present results for the 2-channel Kondo model, the prototype of all non-Fermi liquid
models, and show that there half of the scattering remains inelastic even at the Fermi energy. 
\end{abstract}

\end{frontmatter}


\section{Introduction}
\label{introduction}

One of the major ingredients of mesoscopic physics is 
quantum interference: it leads  to phenomena such as weak localization,
Aharonov-Bohm interference,
universal conductance fluctuations,
or mesoscopic local density of states fluctuations~\cite{Altshuler_review}.
All these  phenomena rely on the phase coherence of the conduction electrons. 
This phase coherence is,  however, 
destroyed through {\em inelastic scattering processes}, where an excitation
is created in the {\em environment}. These inelastic processes suppress  quantum interference 
after the so-called dephasing time, $\tau_\varphi$, also called inelastic
scattering time. 
The excitations created in course of an inelastic scattering process may be phonons, magnons, 
electromagnetic  radiation, or simply electron-hole excitations.

A few years ago Mohanty and Webb measured  the dephasing time
$\tau_\varphi(T)$ carefully down to very low temperatures  through weak localization
experiments, and reported a surprising saturation of it at the lowest 
temperatures~\cite{Mohanty-Webb}.  
These experiments gave rise to many  theoretical speculations:  
 intrinsic dephasing due to electron-electron
 interaction~\cite{Zaikin} as well as   
scattering from two-level systems~\cite{JanZawa,Imry} have been proposed to explain 
the observed saturation, and induced rather violent discussions~\cite{Zaikin,Aleiner,Delft}.
Recently, it has been finally proposed that an apparent saturation could  well
be explained  by  inelastic scattering from magnetic
impurities~\cite{Birge2003,zar_inel}.  

Triggered  by these results of Mohanty and Webb, a number of experimental groups
also revisited the problem of inelastic scattering and dephasing in quantum wires
and disordered metals: A series of experiments have been performed
to study   the non-equilibrium relaxation 
of the energy distribution function in short 
quantum wires~\cite{Saclay}.
These energy relaxation experiments could be well explained in terms of  
the orthodox theory of electron-electron  interaction in one-dimensional
wires~\cite{AAK},  
and/or inelastic scattering mediated by magnetic
impurities~\cite{Zawadowski,Glazman,Grabert,Kroha}.
Parallel to, and partially triggered by these experiments, a systematic study
of the inelastic scattering from 
magnetic impurities has also been carried out recently, where inelastic scattering 
from magnetic impurities at energies down to well below the Kondo scale has also been
studied~\cite{Saminadayar2005,Saminadayar2006,Birge2006}.
 
Theoretically, this strong coupling regime can be reached only 
through a {\em non-perturbative} approach.  Such a method 
to compute the inelastic scattering cross-section has been proposed 
in Ref.~\cite{zar_inel} and further developed in Ref~\cite{Rosch}, 
where it has been shown  that the finite temperature version of the formula introduced in 
Ref.~\cite{zar_inel} describes indeed  the dephasing rate that appears in 
the expression of weak localization in the limit of small concentrations. 
 Except for very low temperatures,  where a small residual 
inelastic scattering is observed~\cite{Saminadayar2006,Birge2006},
these calculations
were in excellent agreement with the experiments, and they clearly showed 
that magnetic impurities in 
concentration 
as small as 1ppm 
can already induce  substantial inelastic scattering.
We have to emphasize though
that experiments on very dirty metals, {\em e.g.}, probably cannot be explained 
in terms of magnetic scattering, and possibly other mechanisms are needed to
account for the dephasing observed at very low temperatures in these systems~\cite{Lin}.

\begin{figure}[tb]
\bc
\includegraphics[width=0.6\columnwidth,clip]{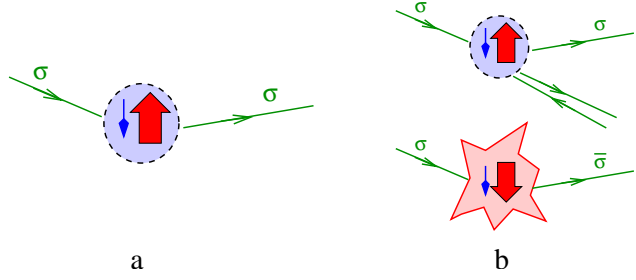}
\ec
\caption{\label{fig:ineastic_elastic}
{Sketch of (a) elastic and (b) inelastic scattering processes.
In case of an inelastic scattering the outgoing 
electron leaves spin- and/or electron-hole excitations behind.}
}
\end{figure}

Here we review the theory of  Ref.~\cite{zar_inel} and show how it can 
be applied to various quantum
impurity problems.  In Ref.~\cite{zar_inel} we formulated the problem of inelastic scattering 
in terms of the  many-body $S$-matrix defined through the overlap of incoming and outgoing
scattering states:
\begin{eqnarray}
\phantom{a}_\textrm{out}\langle f |i\rangle_\textrm{in}
\equiv \phantom{a}_\textrm{in}\langle f
|\hat S|i\rangle_\textrm{in}\;.
 \end{eqnarray}
The incoming and outgoing scattering states, $|i\rangle_\textrm{in}$ and
$|f\rangle_\textrm{out}$,  are asymptotically free, however, they may contain many excitations, 
i.e. they are true many-body states. 
The many-body $T$-matrix is defined as the
 'scattering part' of the $S$-matrix,
$\hat{S}=\hat{I}+i\hat{T}$, 
 with $\hat{I}$  the identity operator. By energy conservation 
\be
_\textrm{in}\langle f |\hat T|i\rangle_\textrm{in}  = 2\pi \; \delta(E_f-E_i)
\; \langle f | {\cal T}|i\rangle \; ,
\label{eq:on_shellT}
\ee
where we introduced  the on-shell $T$-matrix $\langle f |{\cal T}|i\rangle$. 
The results of Ref.~\cite{zar_inel} rely on the simple observation, that 
$\langle {\bf p}\sigma |{\cal T}| {\bf p'}\sigma' \rangle$
determine both the total  ($\sigma_{\rm tot}$) and the 
elastic 
($\sigma_{\rm el}$)
scattering cross sections of the conduction
electrons (or holes) at $T=0$ temperature. 
The total scattering cross section of an electron of momentum 
${\bf p}$ and spin $\sigma$ is given by the optical theorem as
 \be
\sigma_{\rm total}^\sigma = \frac{2}{v_F} {\rm Im} \langle {\bf p} \sigma | {\cal T} | {\bf p} \sigma\rangle
\;, 
\label{eq:sigma_tot}
\ee
where $v_F$ denotes  the Fermi velocity. This expression accounts also for
processes where a single electron scatters into many excited electron states
(see Fig.~\ref{fig:ineastic_elastic}).
In case of elastic scattering, on the other hand, an incoming single electron state is
scattered into an outgoing single electron state, without inducing any 
spin or electron-hole excitation  of the environment. The 
corresponding cross section can be expressed as
\be
\sigma_{\rm el}^\sigma  
= \frac{1}{v_F} \int {\frac{d {\bf p}'}{(2\pi)^3}}
2\pi \; \delta(\xi'-\xi) |\langle {\bf p}' \sigma | {\cal T} | {\bf p}
\sigma\rangle|^2 \;,
\label{eq:sigma_el}
\ee
with $\xi$ the energy of the electron measured from the Fermi surface.
Inelastic scattering processes, schematically illustrated in
Fig.~\ref{fig:ineastic_elastic}, can be defined as scattering processes, which
are {\em not} elastic. Accordingly, the inelastic scattering 
cross section associated with these processes is just 
the {\em difference} of these two cross-sections:
\begin{equation}
\sigma_{\rm inel}^\sigma = \sigma_{\rm total}^\sigma - \sigma_{\rm
  el}^\sigma \; . 
\label{eq:difference}
\end{equation}
This simple formula allows us to compute the inelastic scattering
cross-section in detail.

In a free electron gas  it is convenient  to introduce
angular momentum channels, $L\equiv(l,m)$, and define the scattering states in 
terms of radially propagating states $|{\bf p},\sigma\rangle \to   ||{\bf
  p}|,L,\sigma \rangle$.
In this basis the on-shell $S$ and $T$-matrices become  matrices 
in the quantum numbers $L$, and 
they depend only on the energy $\omega$ of 
the incoming particle~\cite{deph_2}, 
\be
{s}_ {L,\sigma;\; L'\sigma'}(\omega) = \delta_{L,L'}\delta_{\sigma,\sigma'}  + i \; {t}_ {L,\sigma;\; L'\sigma'}(\omega)
\;. 
\ee
By unitarity, the eigenvalues $s_\lambda$ of the matrix $s_ {L,\sigma;\; L'\sigma'}$  
must all be within the complex unit circle for any $\omega$, and they 
are directly related to the inelastic scattering cross section. 
In case of $s$-scattering and spin-conservation, e.g., $s_ {L,\sigma;\; L'\sigma'}$ becomes a simple 
number, $s(\omega)= 1 + i\; t(\omega)$,  
and the inelastic scattering cross section  can be expressed as 
\be
\sigma_{\rm inel}(\omega)= \frac{\pi}{p_F^2}\;(1-|{s}(\omega)|^2)
= \frac{\pi}{p_F^2}\;(2 \;{\rm Im}\; t(\omega)-|t(\omega)|^2)\;,
\label{eq:sigma_inel}
\ee
where we assumed  free electrons of dispersion $\xi={\bf p}^2/2m-\mu$
with  a Fermi energy $\mu$ and a corresponding Fermi momentum $p_F$.
Eq.~(\ref{eq:sigma_inel}) implies that 
 the scattering becomes {\em totally elastic} whenever ${s}(\omega)$ is on
the unit circle, and it is {\em maximally inelastic} if the corresponding 
single particle matrix element of the $S$-matrix vanishes.  The former situation occurs at Fermi
liquid fixed points, while the latter case is realized, e.g., in case of the
two-channel or the two-impurity Kondo models.
The total scattering cross section, on the other hand, is related 
to the real part of ${s}(\omega)$ as
\be
\sigma_{\rm tot}(\omega)= \frac{2\pi}{p_F^2}\;(1- {\rm Re}\{{s}(\omega)\})
= \frac{2\pi}{p_F^2}\;{\rm Im}\{{t}(\omega)\}\;.
\ee
It is easy to generalize this result  to the case of many
scattering channels, and one finds that inelastic scattering can take place 
only  if some of the eigenvalues 
of $s_ {L,\sigma;\; L'\sigma'}$ are not on the unit circle~\cite{singular}.

To determine $t_{L\sigma,L'\sigma'}$ we need to compute 
the  matrix element  
$\langle {\bf p} \sigma | {\cal T} | {\bf p}' \sigma'\rangle$, 
that we first relate to the conduction electrons' Green function 
through the so-called reduction  formula~\cite{deph_2,Itzykson80},
\begin{eqnarray}
\label{eq:T-matrix}
\langle  {\bf p},\sigma| {\cal T} | {\bf p}'  \sigma'\rangle = 
  -  \; [G^0]^{-1}_{\pm{\bf p},\pm\sigma}(\xi)\; 
G_{\pm\vp \;\pm\sigma, \pm  \vp'\;\pm \sigma'} (\xi) \; 
[G^0]^{-1}_{\pm {\bf p}'\pm \sigma'}(\xi) 
\;.
\end{eqnarray}
Here the $\pm$ signs correspond to electron and hole states
with $\xi > 0$ and  $\xi < 0$ and of energy $E=|\xi|$, 
$G_0$ denotes the free electron Green's
function, and $G$ the full many-body time-ordered electron Green's function. 
By Eq.~(\ref{eq:T-matrix}) the positive frequency part of the Green's 
function describes the scattering of electrons, while the negative 
frequency part that of holes. In case of a degenerate vacuum state 
one must average over the various vacuum states  
in Eq.~(\ref{eq:T-matrix}) \cite{singular,Rosch}.

According to Eqs.~(\ref{eq:sigma_tot}), (\ref{eq:sigma_el}), 
(\ref{eq:difference}),  and  (\ref{eq:T-matrix}),  
to compute the inelastic and elastic  scattering cross-sections, 
we only need to evaluate the self-energy of the conduction
electron's Green function, many cases referred to as the $T$-matrix. 
This can be done either analytically using, e.g., perturbative methods, 
or numerically,  by relating the self-energy to some local
correlation function, and computing the latter  by Wilson's numerical 
renormalization group (NRG) method~\cite{NRG_ref}. The latter approach 
enables us to compute both the imaginary and real parts of the matrix
$t(\omega)$, and we can thus also determine the complex eigenvalue 
$s(\omega)$.

\section{Inelastic scattering in the Kondo model}
\label{sec:inel1CK}

Let us first apply this formalism to study the inelastic
scattering in the 
single-channel Kondo model  defined by the Hamiltonian
\be
H=
\sum_{{\bf p}, \sigma}  
\xi_{\bf p}\; a^{\dagger}_{{\bf p} \sigma}a_{{\bf p} \sigma}
+ \frac J2  \vec{S} 
\sum_{{\bf p},{\bf p}' \atop \sigma\sigma'}
a^{\dagger}_{{\bf p}\sigma} {\vec{\sigma}}_{\sigma\sigma'} 
a_{{\bf p}',\sigma'} \;.
\label{Eq:Kondo}
\ee
Here  $a^\dagger_{{\bf p}\sigma}$ creates a conduction electron with
momentum  $\bf p$, spin $\sigma$, and $S=1/2$ is the impurity
spin. The $T$-matrix of the Kondo model can be related to
the Green's function of the so-called composite Fermion operator, 
$F_{\sigma} \equiv \sum_{\sigma', {\bf p}} 
{\vec S} \cdot {\vec \sigma}_{\sigma\sigma'} a_{{\bf p}
  \sigma'}$~\cite{Costi}, whose spectral function  
can then be computed using NRG~\cite{NRG_ref}. 

\begin{figure}
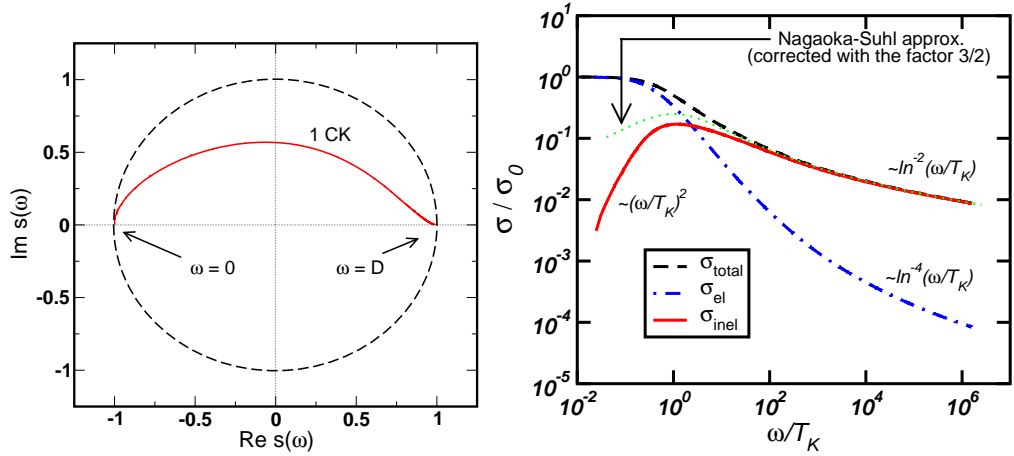

\bc
\includegraphics[width=0.45\columnwidth,clip]{proc_1CK_Sflow.eps} 
\includegraphics[width=0.5\columnwidth,clip]{proc_nagaoka.eps}
\ec
\caption{\label{fig:Sflow}
Left: Renormalization group flow of the eigenvalue of the single 
particle $S$-matrix for the single-channel Kondo
model. Right: Energy-dependence of the elastic-, total- and inelastic scattering rates 
for the Kondo model.}
\end{figure}

Before presenting our numerical results, let us briefly discuss what we can
learn about the inelastic scattering cross-section from analytical approaches. 
The $\omega\gg T_K$ regime is  accessible by perturbation theory. 
Summing up the leading logarithmic diagrams we find
\be
t(\omega\gg T_K) \approx   i \;({\pi^2}/2)
S(S+1) / {\ln^2(\omega/T_K)}\;,
\ee
where $T_K \sim E_F e^{-1/J\varrho }$ is the Kondo temperature, with $E_F$ 
the Fermi energy and $\varrho$ the density of states at the 
Fermi energy for one spin direction~\cite{Hewson}. 
To leading logarithmic order, the scattering is completely {\em inelastic} 
\be
\sigma_{\rm inel}(\omega\gg T_K) 
\approx \sigma_{\rm tot}(\omega) \approx \frac{\pi^3}{p_F^2} \;
S(S+1)  \frac 1 {\ln^2(\omega/T_K)}\;,
\label{sigma_inel_asymp}
\ee
since the elastic contribution only increases as $\sigma_{\rm el}\sim  |t(\omega)|^2$, 
\be
\sigma_{\rm el}(\omega\gg T_K) \approx \frac{\pi^5}{4\;p_F^2} \;
S^2(S+1)^2  \frac 1 {\ln^4(\omega/T_K)}\;. 
\label{sigma_el_asymp}
\ee
This  very surprising result contradicts conventional wisdom,
which tries to associate inelastic scattering with spin-flip scattering. 
It can be  explained in the following way~\cite{Garst}: At high energies, 
incoming electrons are scattered 
by  the impurity spin fluctuations. These fluctuations can absorb an energy of 
the order of $\sim   T_K$, and therefore the energy of the incoming electron
is not conserved even in leading order, but it typically changes by a tiny 
amount, $ \delta\omega \sim T_K$. In earlier approaches, this tiny energy 
transfer has been neglected, and the spin-diagonal scattering has been
incorrectly identified as an elastic process.

We can also relate the cross sections above to scattering rates. 
Assuming a finite but small concentration $n_{\rm imp}$ of magnetic impurities, the 
conduction electrons' lifetime can be expressed as 
$1/\tau(\omega) = n_{\rm imp}\; v_F \; \sigma_{\rm imp}(\omega)$. This relation can be
used to define the inelastic scattering rate too as
\be 
\frac 1 {\tau_{\rm inel}} \equiv n_{\rm imp}\; v_F \; \sigma_{\rm inel}(\omega)
\approx n_{\rm imp}   \frac {  \pi  S(S+1) } {2\;\varrho
  \;\ln^2(\omega/T_K)}\;, \phantom{nn} (\omega\gg T_K).
\ee  
Note that the latter asymptotic expression  is a factor 3/2 larger than the Nagaoka-Suhl
formula, which takes into account only spin-flip processes~\cite{Nagaoka}.

For energies $|\omega| < T_K$ perturbation theory breaks down, and it is more
appropriate to use Nozi\`eres' Fermi liquid theory,  according to which
scattering at Fermi energy scattering is completely elastic, and is described
through simple phase shifts \cite{Nozieres}, $t_\sigma(\omega=0^+)= 2 \sin \delta_\sigma\;e^{i\delta_\sigma}$.
Here we allowed for different phase shifts in the spin up and spin down
channels. This Fermi liquid expression yields
\be
\sigma_{\rm tot,\sigma}(\omega\to 0) = \frac {4\pi}{p_F^2}
\sin^2(\delta_\sigma)\;,\phantom{nnnnnnnn} \sigma_{\rm inel, \sigma}(\omega\to 0) = 0\;.
\ee
The maximum total scattering cross section is  reached in the unitary limit,
$\delta_\sigma=\pi/2$, while the inelastic scattering cross-section always vanishes at the
Fermi energy. Perturbation theory around the Fermi liquid fixed point predicts 
$\sigma_{\rm inel, \sigma}(\omega\to 0) \propto  (\omega/T_K)^2$ \cite{zar_inel,NozieresII}.

The analytical calculations above can only capture the physics in the limit of
very large and very small frequencies, and for energies $\omega\sim T_K$  
we need to use more sophisticated methods such as NRG.
In Fig.~\ref{fig:Sflow}  we show the evolution  of the eigenvalue of the
$s(\omega)$. In the limit $T_K/E_F\ll 1$ this becomes a universal function, 
${s}(\omega) = {s}(\omega/T_K)$.  In the single-channel Kondo model  $|{s}(\omega)|\to 1$ for both very large
and very small frequencies, and thus  scattering becomes  
completely elastic both limits.  The reasons are different: At large 
energies conduction electrons 
do not interact with the impurity spin efficiently. 
At very small energies, on the other hand, the impurity's spin is screened and
disappears from the problem~\cite{Nozieres}.
The maximum inelastic scattering is reached when the eigenvalue ${s}(\omega)$ is  closest to the
origin, i.e., at energies in the range of the Kondo temperature,
$\omega\approx T_K$.

\begin{figure}
\bc
\includegraphics[width=8cm,clip]{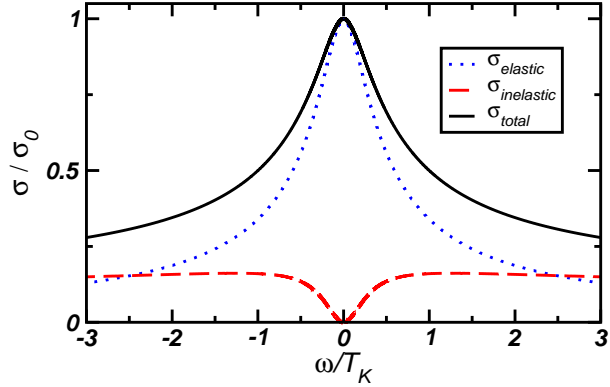}
\ec
\caption{
\label{fig:Inel_1CK}
Inelastic, elastic, and
  total scattering 
  rates
  for the single-channel Kondo model
  in units of $\sigma_0 = 4\pi/p_F^2$,
  as a function of the 
  incoming electron's energy.   
}
\end{figure}

The total, elastic and inelastic scattering cross sections of an electron
are shown in Fig.\ref{fig:Inel_1CK}. As expected, the
inelastic amplitude always vanishes at the Fermi level, 
and at energies well above $T_K$ 
both the elastic and  the inelastic scattering cross-sections  follow the 
the analytical expressions, Eqs.~(\ref{sigma_inel_asymp}) and
(\ref{sigma_el_asymp}). For energies  $|\omega|\ll T_K$ we recover 
the quadratically  vanishing inelastic rate expected from Fermi liquid
theory~\cite{NozieresII}, 
but the $\sigma_{\rm inel}\sim\omega^2$ regime appears only at
energies well below the Kondo temperature, $\omega<0.05\; T_K$ 
(see Fig.\ref{fig:Inel_1CK}).
 For $|\omega|\gg T_K$ the usual Nagaoka-Suhl
expression  describes the 
inelastic scattering rather well apart from the incorrect overall 
pre-factor 3/2 discussed before, but it
starts to deviate strongly from the numerically exact curve at approximately
$10\;T_K$, and it completely fails below the Kondo temperature $T_K$.

Interesting features appear also at intermediate temperatures. 
The inelastic
scattering rate is roughly linear between $0.05\;T_K<\omega<0.5\;T_K$,
and a  broad plateau appears above the Kondo scale, where the 
energy-dependence of the inelastic scattering rate turns out to be 
extremely weak.
Even though our calculation is done at  $T=0$ temperature, 
$\sigma_{\rm inel}(T,\omega=0)$ is expected to behave very similarly 
to $\sigma_{\rm inel}(T=0,\omega)$.  Thus both features 
are perfectly consistent with several  
experiments~\cite{Mohanty2003,Saminadayar2005,Birge2003},
and, somewhat surprisingly, our results fit  the
experimentally measured temperature-dependence of $1/\tau_\Phi$
excellently~\cite{Saminadayar2005}. This is beyond expectations, 
since realistic magnetic impurities have a complicated 
$d$-level structure, 
and our $T=0$ temperature results provide just approximations 
for the dephasing rate, and  one should use the finite temperature 
expression of the dephasing rate, computed in Ref.~\cite{Rosch}.

\section{Anderson model}
\label{sec:inelAnderson}

Let us next discuss the Anderson model,  defined by the Hamiltonian, 
 \begin{eqnarray}
 H=\sum_{\vp\sigma}\epsilon(\vp) a^{\dagger}_{\vp
   \sigma}a^{\phantom{\dagger}}_{\vp\sigma}
+\epsilon_d\sum_{\sigma}d^{\dagger}_\sigma d^{\phantom{\dagger}}_\sigma
+
Ud^{\dagger}_{\uparrow}d^{\phantom{\dagger}}_{\uparrow}d^{\dagger}_{\downarrow} d^{\phantom{\dagger}}_{\downarrow}
+ V \sum_{\sigma,\vp } \left( c^{\dagger}_{\vp \sigma}  d^{\phantom{\dagger}}_\sigma+\textrm{h.c.}\right)\;. 
\nonumber
 \end{eqnarray}
Here  $d_\sigma$ denotes a local $d$-level's annihilation operator, 
$U$ is the on-site Coulomb repulsion, and 
the conduction band and the local electronic level are hybridized by $V$. 
The Anderson model is the most elementary Hamiltonian that describes
local moment  formation on a localized $d$-orbital, and in fact, 
the Kondo Hamiltonian can be obtained from it  in the 
limit $\Delta\ll U+\epsilon_d, |\epsilon_d|$, with $\Delta = \pi \varrho V^2$
the width of the resonance~\cite{Hewson}.   

As first discussed by Langreth~\cite{langreth}, 
the $T$-matrix for the Anderson model can be related to the $d$-level's 
Green's function as~\cite{zar_inel,Hewson}
\be
{\rm Im} \{ {\cal T}_{\sigma}(\omega)   \} =  \pi  {V^2}
\varrho_{d,\tau \sigma}(\omega)\;, 
\phantom{nn}
{\rm Re}\{ {\cal T}_{\sigma}(\omega)\} =  - \tau \;
  {V^2} \int d\omega'\; \frac{\varrho_{d,\tau \sigma}(\omega)}{\omega-\omega'}
\;. 
\nonumber
\ee
Here $\tau={\rm sgn}\; \omega$, and  $\varrho_{d,\sigma}(\omega)$ is the spectral
function of the $d$-Fermion, which we have computed using NRG \cite{deph_2}.

\begin{figure}
\begin{center}
\includegraphics[width=0.6\columnwidth,clip]{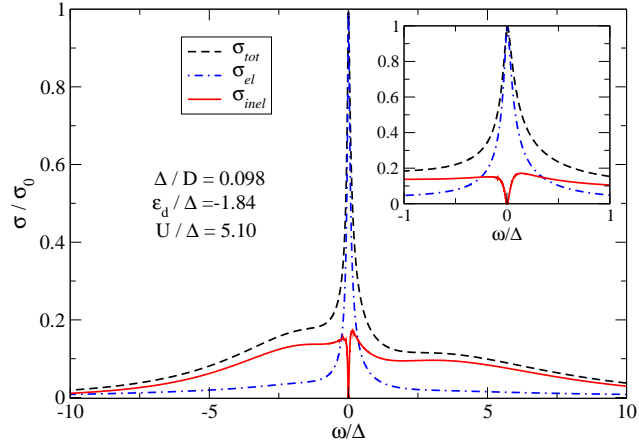}
\end{center}
\caption{
Elastic, inelastic and total scattering cross section
for the asymmetric Anderson model. The low-energy part of the curves is very similar to the one 
obtained for the Kondo model.
}
\label{anderson_asym}
\vspace{0.5cm}
\end{figure}

The full frequency-dependence of the various scattering cross-sections
obtained for the asymmetrical Anderson model with interaction strength 
$U/\Delta = 5.1$ is shown in Fig.~\ref{anderson_asym}. At this value of $U$
one can already observe the Hubbard side-peaks in the total scattering cross
section at energies 
$\omega\approx \epsilon_d$ and $\omega\approx U + \epsilon_d$,  and a distinct 
Kondo resonance appears at $\omega\approx 0$ too. 
The scattering rates in the region $\omega\approx 0$ are    
strikingly similar to the ones we  obtained for the Kondo model, 
and remarkably, both the quasi-linear regime of  $\sigma_{\rm inel}$ and the
plateau  are already  present for these moderate values of $U/\Delta$. 
This is not very surprising since, as stated before,  
the Kondo model is just the effective model of the Anderson 
 model in the limit of large $U/\Delta$ and $\omega \ll U$.
For even larger values of $U/\Delta$ and intermediate energies, 
$T_K\ll\omega\ll U$,  the elastic and   inelastic contributions  
follow very nicely the  asymptotic behavior found for the Kondo model, and
scale as $\sim 1/\ln^4(\omega/T_K)$ and $\sim 1/\ln^2(\omega/T_K)$,
respectively. New features compared to the Kondo model are the 
Hubbard peaks that correspond almost entirely to  inelastic scattering.

\section{Inelastic scattering in the two-channel Kondo model}
\label{sec:inel2CK}

So far, we considered scattering from  Fermi liquid models only. Let us now 
discuss the 
 two-channel Kondo model, the prototype of all non-Fermi liquid impurity 
models~\cite{Cox}. This is defined by a Hamiltonian similar to 
Eq.~(\ref{Eq:Kondo}) excepting that now there is two 'channels' of conduction 
electrons, $\alpha=1,2$ that are coupled to the impurity spin with couplings $J_\alpha$,  
\be
H=
\sum_{\alpha=1,2} \sum_{{\bf p}, \sigma}  
\xi_{\bf p}\; a^{\dagger}_{{\bf p} \sigma,\alpha}a_{{\bf p} \sigma,\alpha}
+ \sum_{\alpha=1,2}   \frac {J_\alpha}2  
\vec{S} 
\sum_{{\bf p},{\bf p}' \atop \sigma\sigma'}
a^{\dagger}_{{\bf p},\alpha} {\vec{\sigma}}_{\sigma\sigma'} a_{{\bf p}',\alpha} 
\;.
\label{Eq:2CK}
\ee
In the channel-symmetric case, $J_1 = J_2$ the
two  conduction electron channels compete to screen the impurity
spin independently, which is therefore never completely screened. 
This competition  
leads to the  formation of a strongly correlated state which cannot be
described by Nozi\`eres' Fermi liquid theory, and is characterized by 
a non-zero residual entropy, the
logarithmic divergence of the impurity susceptibility, and the power
law behavior of transport properties with fractional exponents \cite{Cox}. 
Any infinitesimal  asymmetry in the couplings
$\Delta=(J_1-J_2)/(J_1+J_2)$ leads to
the appearance of another low-temperature
energy scale $T^*\propto \Delta^2\; T_K/$ at which the system crosses over to 
a Fermi liquid behavior: Electrons
being more strongly coupled to the impurity  form a usual Kondo
singlet with the impurity spin, while the other electron channel becomes
completely  decoupled from the spin.

For $J_1 = J_2$, no Fermi-liquid
relations are available. There exists, however, 
an exact theorem due to Maldacena and Ludwig, according to which,
at the two-channel Kondo fixed point, the   
single-particle elements of the $S$-matrix identically vanish
for $\omega\to 0$: $s_{2CK}(\omega\to 0)= 0$~\cite{Maldacena}.
As a consequence,  $t_{2CK} (\omega=0)= -i \;.$
This relation leads to the surprising result 
that exactly half of the scattering is inelastic 
at the Fermi energy, while the other half of it is inelastic:
\be
\sigma_{\rm inel}^{\rm 2CK}(\omega=0)=\sigma_{\rm el}^{\rm 2CK}(\omega=0)=
{\sigma_{\rm tot}^{\rm 2CK}(\omega=0)}/2\;.
\ee
This counter-intuitive result can be understood as follows: The 
vanishing of the  single particle $S$-matrix indicates that
an incoming electron {\em cannot} be detected as one electron after 
the scattering event, and it ``decays'' into infinitely many electron-hole
pairs. To get such a ``decay'', however, the scattering process must have an
elastic part which interferes destructively with the unscattered
direct wave, and cancels exactly the outgoing single particle
amplitude in the $s$-channel.

\begin{figure}
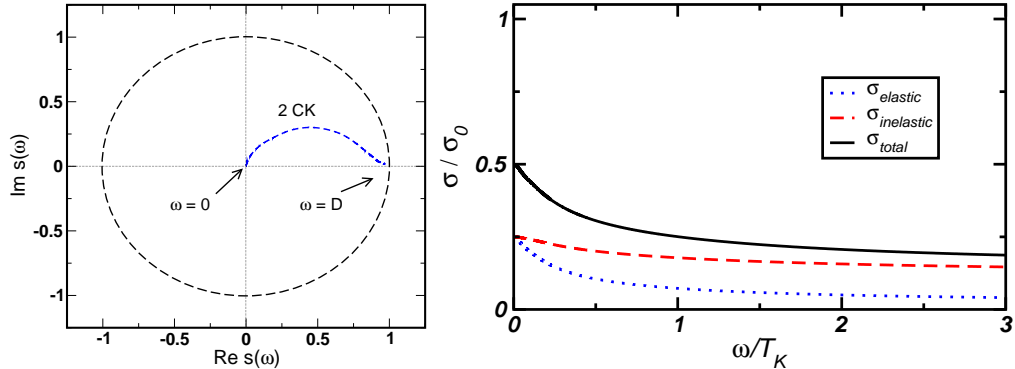

\bc
\includegraphics[width=0.4\columnwidth,clip]{proc_2CK_Sflow.eps}
\includegraphics[width=0.55\columnwidth,clip]{2ck_v3.eps}
\ec
\caption{\label{fig:2CK_inel}
Left: Renormalization group flow of the eigenvalue of the single 
particle $S$-matrix for the two-channel Kondo
model.
Right: Energy-dependence of
  elastic and inelastic scattering rates 
  for the two-channel Kondo model
 in units of
  $\sigma_0 = 4\pi/p_F^2$, at $T=0$.
}
\end{figure}

The evolution of $s(\omega)$ and the inelastic scattering rates for the two-channel Kondo model are 
shown in Fig.~\ref{fig:2CK_inel} as a function of the energy of the incoming
particle. 
In the channel-symmetric case inelastic  processes are allowed  
even at $\omega=0$, which is a clear signature of the
non-Fermi liquid behavior. 
The non-Fermi liquid nature
is also reflected in the $\sim\sqrt{\omega}$ singularity of
the scattering cross sections at $\omega=0$.

For $\Delta>0$ the total scattering rate approaches the unitary limit
in channel ``1'' below the Fermi liquid scale $T^*$,  while it goes to 0
for $\Delta<0$. In both cases, the inelastic scattering  freezes out, 
$\sigma_{\rm inel}(\omega)$ shows a dip below $T^*$, and it ultimately scales to 0 as
$$
\sigma_{\rm inel}(\omega)\propto  {\omega^2}/{{T^*}^2}\;.
$$
Remarkably, 
the inelastic scattering cross-section is 
very similar for $\Delta>0$  and  $\Delta<0$, while the 
scattering contributions are dramatically different in these two cases
(see Ref.~\cite{deph_2}). 

\section{Conclusions}

In the present paper we reviewed a recent theory of inelastic scattering from
quantum impurities, and applied it to the single and two-channel Kondo models,
and the Anderson model. 

We showed that in the Kondo model and in the local moment regime of the
Anderson model a broad plateau appears in the $T=0$ temperature
energy-dependent  inelastic scattering rate above $T_K$, while a quasi-linear
regime emerges below $T_K$, in the $T=0$ temperature, both in excellent
agreement with recent experimental observations. 
 We also computed the universal
flow of the eigenvalues  of the $S$-matrix, which is a useful quantity to classify
various types of impurity states~\cite{singular}. 

As an example of a non-Fermi liquid, we also discussed scattering from the
two-channel Kondo model, where half of the scattering remains  inelastic 
even at the Fermi energy, and correspondingly, the eigenvalue of the  
$S$-matrix vanishes. This fragile non-Fermi liquid state is, however,
destroyed, once  a small  channel-symmetry breaking is applied, and  then  
the scattering becomes elastic below a Fermi-liquid scale, $T^\star$ \cite{deph_2}.

Our results   have been  extended  to $T\ne0$ temperatures 
by Micklitz et al. in Ref.~\cite{Rosch}, who showed that 
the dephasing time can indeed be related to the inelastic cross-section 
computed here for $T=0$ temperature. 
The theory presented here has also been applied by Koller {\em et al.} 
to the $S=1$ Kondo model~\cite{Hewson2}. This under-screened model represents 
a singular Fermi liquid~\cite{singular}, where the scattering is elastic at
the Fermi energy, however, it scales to 0 only logarithmically, in contrast to
a Fermi liquid.  

Let us close our conclusions with an important remark.
In a real experiment, the external electromagnetic field couples with a
minimal coupling to the {\em conduction electrons}. As a consequence,  
the Kubo formula is formulated in terms of the conduction electron 
current operators, and the relevant quantity to
determine dephasing  is thus   the inelastic scattering rate 
 of {\em electrons}. This is what we compute here and that has been computed 
 in Ref.~\cite{Rosch}.  Quasiparticles are, on the other hand, typically not minimally 
coupled to the guage field, since they are usually complicated 
objects in terms of conduction electrons. If one defines quasiparticles 
as {\em stable} elementary excitations of the vacuum, as Nozi\`eres 
did \cite{Nozieres}, or as they appear in Bethe Ansatz, then, by definition, 
these quasiparticles do not decay at all at $T=0$ and scatter only elastically \cite{Nozieres}. 
However, excepting for $\omega=0$, a real conduction electron is composed of {\em
 many} such  stable quasiparticles, and it already decays inelastically even at $T=0$ temperature.
In the Kondo model, at the Fermi energy quasiparticle states are just
phase shifted conduction electron states, however, the connection between 
quasiparticles and conduction electrons is not trivial for any finite energy.
Therefore, if one considers inelastic scattering at a finite energy, one must
{\em precisely} specify  
how finite energy quasiparticle states are defined, how they couple to a guage
field, and how a finite energy electronic state is decomposed in terms of
these quasiparticles,. 
In the present framework, we avoid this difficulty by formulating the
problem in terms of electrons.

We are indebted to  L. Saminadayar, C. B\"auerle, J.J. Lin, and A. Rosch 
for valuable discussions. 
This research has been supported by Hungarian grants Nos.
NF061726, D048665, T046303 and  T048782.
L.B. acknowledges the financial support of the Bolyai 
Foundation.




\begin{thebibliography}{00}




\bibitem{Altshuler_review} 
For a review see,{\em e.g.}, B.L. Altshuler, in {\em Les Houches 
Lecture Notes on Mesoscopic Quantum Physics} (edited by A. Akkermans {\em et al.}, 
Elsevier, 1995). 
\bibitem{Mohanty-Webb} 
P. Mohanty, E. M. Q. Jariwala, and R. A. Webb, Phys. Rev. Lett. {\bf 78}, 
3366 (1997).
\bibitem{Zaikin} D. S. Golubev and A. D. Zaikin, Phys. Rev. Lett. {\bf 81},
  1074 (1998); See also the criticism in 
\bibitem{JanZawa} 
A. Zawadowski, J. von Delft, and D. C. Ralph, Phys. Rev. Lett. {\bf 83}, 2632 (1999).
\bibitem{Imry}
Y. Imry, H. Fukuyama and P. Schwab, Europhys. Lett. {\bf 47}, 608 (1999).
\bibitem{Aleiner} 
I. L. Aleiner, B. L. Altshuler and M. E. Gershenson, Waves in Random Media {\bf 9}, 201 (1999).
\bibitem{Delft} J.~von~Delft, cond-mat/0510563; J.~von~Delft, in~{\it Fundamental Problems of Mesoscopic Physics}, edited by I.V. Lerner {\it et al.} (Kluver, London, 2004), p.115-138.
\bibitem{Birge2003}  F. Pierre, A. B. Gougam, A. Anthore, H. Pothier, D. Esteve, and N. O. Birge, Phys. Rev. B {\bf 68}, 085413 (2003).
\bibitem{zar_inel} G. Zar\'and, L. Borda, J. von Delft and N. Andrei,
Phys. Rev. Lett. {\bf 93}, 107204 (2004).
\bibitem{Saclay}H. Pothier, S. Gu\'eron, N. O. Birge, D. Esteve, and
  M. H. Devoret,
Phys. Rev. Lett. {\bf 79}, 3490 (1997).
\bibitem{AAK}B. L. Altshuler, A. G. Aronov, and D. E. Khmelnitskii, 
J. Phys. C {\bf 15}, 7367 (1982).
\bibitem{Zawadowski}
 J. S\'olyom and A. Zawadowski, Z. Phys. {\bf 226}, 116 (1969).
\bibitem{Glazman}
A. Kaminski and L. I. Glazman, Phys. Rev. Lett. {\bf 86}, 2400 (2001).
\bibitem{Grabert} 
G. G\"oppert, Y. M. Galperin, B. L. Altshuler, and H. Grabert,       
Phys. Rev. B {\bf 66}, 195328 (2002).
\bibitem{Kroha} 
J. Kroha and A. Zawadowski, Phys. Rev. Lett. {\bf 88}, 176803 (2002).
\bibitem{Saminadayar2005} 
F. Schopfer, C. Bäuerle, W. Rabaud, and L. Saminadayar 
Phys. Rev. Lett. {\bf 90}, 056801 (2003); 
C. B\"auerle, F. Mallet, F. Schopfer, D. Mailly,
  G. Eska, and L. Saminadayar, Phys. Rev. Lett. {\bf 95}, 266805 (2005).
\bibitem{Saminadayar2006} F. Mallet, J. Ericsson, D. Mailly, S. \"Unl\"ubayir,
  D. Reuter, A. Melnikov, A. D. Wieck, T. Micklitz, A. Rosch, T. A. Costi,
  L. Saminadayar, and C. B\"auerle, Phys. Rev. Lett. {\bf 97}, 226804 (2006).
\bibitem{Birge2006} G. M. Alzoubi and N. O. Birge, 
Phys. Rev. Lett. {\bf 97}, 226803 (2006).
\bibitem{Rosch} T. Micklitz, A. Altland, T. A. Costi, A. Rosch,
Phys. Rev. Lett. {\bf 96}, 226601 (2006).
\bibitem{Lin} J. J. Lin, private communication.
\bibitem{deph_2} 
L. Borda, L. Fritz, N. Andrei, and  G. Zar\'and, submitted to 
Phys. Rev. B.
\bibitem{Itzykson80} 
C. Itzykson and J. B. Zuber, {\em Quantum Field The\-o\-ry} (McGraw-Hill, 1985). 
\bibitem{Nozieres} P.~Nozi\`eres, J. Low Temp. Phys. {\bf 17}, 31 (1974).  
\bibitem{Garst} M. Garst, P. W\"olfle, L. Borda, J. von Delft, L. I. Glazman,
Phys. Rev. B {\bf 72}, 205125 (2005).
\bibitem{NozieresII}In Ref.~\cite{Nozieres} Nozi\`eres considers stable 
quasiparticles rather then electrons, which do not decay at $T=0$.
\bibitem{Mohanty2003}P. Mohanty and R. A. Webb,
Phys. Rev. Lett. {\bf 91}, 066604 (2003).
\bibitem{langreth} 
 D. C. Langreth,  Phys. Rev. {\bf 150}, 516 (1966). 

\bibitem{Cox} D.L. Cox, A. Zawadowski, Adv. Phys. {\bf 47} 599 (1998). 

\bibitem{Hewson} For a review see A.C Hewson, 
{\em The Kondo Problem to Heavy Fermions}, Cambridge University Press (1993).
\bibitem{Costi}
 T. A. Costi,  Phys. Rev. Lett. {\bf 85}, 1504 (2000).
\bibitem{singular} P. Mehta, N. Andrei, P. Coleman, L. Borda, G.
  Zar\'and, Phys. Rev. B {\bf 72}, 014430 (2005).
\bibitem{Maldacena} J. M. Maldacena and A. W. W. Ludwig, Nucl. Phys. {\bf B506}, 565 (1997).
\bibitem{Nagaoka}Y. Nagaoka
Phys. Rev. {\bf 138}, A1112 (1965); H. Suhl, Phys. Rev {\bf 138}, A515 (1965).
\bibitem{NRG_ref} 
K.G. Wilson, Rev. Mod. Phys. {\bf 47}, 773 (1975);
for a recent review see R. Bulla, T. A. Costi, T. Pruschke,
cond-mat/0701105 (2007).

\bibitem{Hewson2} W. Koller, A. C. Hewson, D. Meyer, Phys. Rev. B {\bf 72}, 045117 (2005).

\end{thebibliography}
\end{document}